\documentclass[reviewcopy]{elsarticle}

\usepackage[reviewcopy]{adndt}
\usepackage{longtable}

\usepackage{amsmath}

\usepackage{amssymb}

\usepackage{lscape}

\biboptions{square,sort&compress}
\bibpunct[]{[}{]}{,}{n}{}{;}
\begin{document}

\begin{frontmatter}

\journal{Atomic Data and Nuclear Data Tables}


\title{Magnetic-dipole-to-electric-quadrupole cross-susceptibilities for relativistic hydrogenlike atoms in some low-lying discrete energy eigenstates}

  \author{Patrycja Stefa{\'n}ska\corref{cor1}}
  \ead{pstefanska@mif.pg.gda.pl}

  \cortext[cor1]{Corresponding author.}

  \address{Atomic Physics Division,
Department of Atomic, Molecular and Optical Physics, 
Faculty of Applied Physics and Mathematics, \\
Gda{\'n}sk University of Technology, 
Narutowicza 11/12, 80--233 Gda{\'n}sk, Poland}

\date{April 18, 2016} 


\begin{abstract}  
In this paper we present tabulated data for magnetic-dipole-to-electric-quadrupole cross-susceptibilities ($\chi_{\textrm{M}1 \to \textrm{E}2}$) for Dirac one-electron atoms with a pointlike, spinless and motionless nucleus of charge $Ze$. Numerical values of this susceptibility for the hydrogen atom ($Z=1$) and for hydrogenic ions with $2 \leqslant Z \leqslant 137$ are computed from the general analytical formula, recently derived by us [P. Stefa{\'n}ska, Phys. Rev. A 93 (2016) 022504], valid for an arbitrary discrete energy eigenstate.  In this work we provide 30 tables with the values of  $\chi_{\textrm{M}1 \to \textrm{E}2}$  for the ground state, and also for the first, the second and the third set of excited states (i.e.: 2s$_{1/2}$, 2p$_{1/2}$, 2p$_{3/2}$, 3s$_{1/2}$, 3p$_{1/2}$, 3p$_{3/2}$, 3d$_{3/2}$, 3d$_{5/2}$, 4s$_{1/2}$, 4p$_{1/2}$, 4p$_{3/2}$, 4d$_{3/2}$, 4d$_{5/2}$, 4f$_{5/2}$ and 4f$_{7/2}$) of the relativistic hydrogenlike atoms.  The value of the inverse of the fine-structure constant used in the calculations is $\alpha^{-1}=137.035 \: 999 \: 139$, and was taken from CODATA 2014.\\
\end{abstract}

\end{frontmatter}


\begin{center}
\large{\textbf{Published as: At. Data Nucl. Data Tables 113 (2017) 316--349}
\\*[1ex]
\textbf{doi: 10.1016/j.adt.2016.04.004}\\*[5ex]}
\end{center}



\newpage

\tableofcontents
\listofDtables
\listofDfigures
\vskip5pc


\section{Introduction}

Theoretical investigation of atomic and molecular electromagnetic susceptibilities has been very commonly reported from early decades of the last century \cite{Vlec32}. Most often, numerical computation of such quantities for many-electron atoms and molecules, as well as analytical calculations for simple systems, can be compared with experimental data, and sometimes they appear to be very helpful in the design of other experiments or improving those already used. Thus, in this work we provide theoretical study of the magnetic-dipole-to-electric-quadrupole cross-susceptibility for hydrogenlike atoms with a pointlike, spinless and infinitely heavy nucleus of charge $Ze$. 

The phenomenon of inducing electric quadrupole moment $\mathcal{Q}$ in such systems by a perturbing uniform (dipole) magnetic field of a strength $B$ was pointed out already in the mid-1950s by Coulson and Stephen \cite{Coul56}. Analytical derivation of the expression for this moment was reported also later by Potekhin and Turbiner \cite{Turb87, Pote01}. However, in all these articles, calculations have been performed within the framework of the \emph{nonrelativistic} quantum mechanics. First fully \emph{relativistic} determination of that quantity has been carried out by us in 2012 \cite{Szmy12}. Using the perturbation theory combined with the technique based on the Sturmian expansion of the generalized Dirac--Coulomb Green function \cite{Szmy97}, we have successfully derived the analytical expression for $\mathcal{Q}^{(1)}(B)$ (the first-order correction to the electric quadrupole moment as a function of the field strength $B$) induced in the \emph{ground} state of Dirac one-electron atom (for other applications of this powerful technique to dipole electromagnetic properties of such system, see Refs. \cite{Szmy02b,Szmy04,Szmy02a,Miel06,Szmy11,Stef12,Szmy14}).

Very recently, Szmytkowski and {\L}ukasik considered analytically \cite{Szmy16} and numerically \cite{Szmy16b} some families of \emph{multipole} electric and magnetic susceptibilities for the ground state of relativistic hydrogenlike atoms. In particular, they pointed out that $2^{\lambda}$-pole electric moments induced in the system by static $2^{L}$-pole magnetic field are closely related to $2^{L}$-pole magnetic moments induced in the atom by static $2^{\lambda}$-pole electric field. To be precise, they indicated that the values of the magnetic-to-electric ($\chi_{\textrm{M}L \to \textrm{E}\lambda}$) and electric-to-magnetic ($\alpha_{\textrm{E}\lambda \to \textrm{M}L}$) cross-susceptibilities are \emph{equal} for ions with the charge numbers from the whole range $1 \leqslant Z \leqslant 137$. This suggests that theoretical knowledge of such kind of quantities -- in particular, the $\chi_{\textrm{M}1 \to \textrm{E}2}$ susceptibility -- may be doubly useful for experimentalists: those who would like to measure one of these atomic parameters but are unable to do that, may try to conduct another experiment which allows to determine the second of them.

Some time ago, we have discussed the aforementioned physical quantity in terms of another generalization. In \cite{Stef16} we derived  analytically the expression for the magnetic-dipole-to-electric-quadrupole cross-susceptibility ($\chi_{\textrm{M}1 \to \textrm{E}2}$) for the atom being in an \emph{arbitrary} discrete energy eigenstate, characterized by the set of quantum numbers $\{n_{r}, \kappa, \mu \}$, in which $n_{r}$ denotes the radial quantum number, the Dirac quantum number $\kappa$ is an integer different form zero, whereas $\mu=-|\kappa|+\frac{1}{2}, -|\kappa|+\frac{3}{2}, \ldots, |\kappa|-\frac{1}{2}$ is the magnetic quantum number. The calculations for such general case were based on the Sturmian expansion of the generalized Dirac--Coulomb Green function \cite{Szmy97} (for other applications of this method to another magnetic dipole properties of excited atomic states, see Refs. \cite{Stef15,Stef16b}). In view of the definition below:
\begin{equation}
\mathcal{Q}^{(1)}=\textrm{sgn}(\mu)  (4\pi \epsilon_0) c \: \chi_{\textrm{M}1 \to \textrm{E}2} B,
\label{eq:1}
\end{equation}
where $c$ is the speed of light, while $\mathcal{Q}^{(1)} \equiv \mathcal{Q}_{20}^{(1)}$ is the only one of the spherical components $\mathcal{Q}_{2M}^{(1)}$ ($M=0,\pm1,\pm2$) of the quadrupole electric moment tensor which can be induced in the atom by the perturbing magnetic field $\boldsymbol{B}=B\boldsymbol{n}_z$ (for details, see Ref.\ \cite{Stef16}),  the formula for the atomic parameter under study can be written as follows:
\begin{eqnarray}
\chi_{\textrm{M}1 \to \textrm{E}2}(n_{r},\kappa,\mu)
&=&
\frac{\alpha a_0^4}{Z^4}  
\frac{|\mu|}{64(4\kappa^2-1)^2} 
\Bigg\{
\Theta_{n_{r}\kappa \mu}^{(\texttt{I})}
+\sum_{\substack{\kappa'=-\infty \\ (\kappa' \neq 0)}}^{\infty} 
\frac{\eta_{\kappa \mu}^{(+)}\delta_{\kappa',-\kappa+1}+\eta_{\kappa \mu}^{(-)}\delta_{\kappa',-\kappa-1}}{N_{n_{r}\kappa}+\kappa'} 
\nonumber \\
&& \times
\Bigg[
\Theta_{n_{r}\kappa}^{(\texttt{II})}
+\frac{3n_{r}!N_{n_{r}\kappa}^3\Gamma(n_{r}+2\gamma_{\kappa}+1)}{(N_{n_{r}\kappa}-\kappa)(\gamma_{\kappa'}-\gamma_{\kappa}-n_{r}+1)\Gamma(2\gamma_{\kappa'}+1)} 
\sum_{k=0}^{n_{r}}\sum_{p=0}^{n_{r}} \mathcal{X}(k) \mathcal{Y}(p) 
\nonumber \\
&& \quad \times 
 {}_3F_2 \left(\begin{array}{c} 
\gamma_{\kappa'}-\gamma_{\kappa}-k-2,\:
\gamma_{\kappa'}-\gamma_{\kappa}-p-1,\: 
\gamma_{\kappa'}-\gamma_{\kappa}-n_{r}+1 \\
\gamma_{\kappa'}-\gamma_{\kappa}-n_{r}+2,\:
2\gamma_{\kappa'}+1
\end{array}
;1 
\right)
\Bigg]
\Bigg\}.
\label{eq:2}
\end{eqnarray}
Here $\alpha$ is the Sommerfeld's fine structure constant, $a_0$ denotes, as usually, the Bohr radius, $\Gamma(\zeta)$ is the Euler's gamma function and ${}_3F_2$ is the generalized hypergeometric function, while
\begin{equation}
\eta_{\kappa \mu}^{(\pm)}=\frac{2\kappa \pm 1}{2\kappa \mp 3}\left[(2\kappa \mp 1)^2-4\mu^2\right],
\label{eq:3}
\end{equation} 
\begin{eqnarray}
\Theta_{n_{r}\kappa \mu}^{(\texttt{I})}=32 \left(\alpha Z\right)^2\kappa^2(4\kappa^2-12\mu^2-1) 
\left\{(5n_{r}^2+10n_{r} \gamma_{\kappa}+4\gamma_{\kappa}^2+1)N_{n_{r}\kappa}^2-\kappa(n_{r}+\gamma_{\kappa})\left[2\kappa(n_{r}+\gamma_{\kappa})+3N_{n_{r} \kappa}\right] \right.\},
\label{eq:4}
\end{eqnarray}
\begin{eqnarray}
\Theta_{n_{r}\kappa}^{(\texttt{II})}=6(N_{n_{r}\kappa}-\kappa)N_{n_{r}\kappa}^2 \left\{
4\kappa (n_{r}+\gamma_{\kappa})^2\left[7(n_{r}+\gamma_{\kappa})^2-3\gamma_{\kappa}^2+5\right]+N_{n_{r}\kappa}\left[63(n_{r}+\gamma_{\kappa})^4+70(n_{r}+\gamma_{\kappa})^3\right. \right.
\nonumber \\
\left. \left. -21(2\gamma_{\kappa}^2-5)(n_{r}+\gamma_{\kappa})^2-10(3\gamma_{\kappa}^2-5)(n_{r}+\gamma_{\kappa})+3(\gamma_{\kappa}^2-1)(\gamma_{\kappa}^2-4) \right] 
\right\},
\label{eq:5}
\end{eqnarray}
\begin{equation}
\mathcal{X}(k)=\frac{(-)^{k}}{k!(n_{r}-k)!} \frac{\Gamma(\gamma_{\kappa}+\gamma_{\kappa'}+k+3)}{\Gamma(k+2\gamma_{\kappa}+1)}\left[C_{k}^{(1)}(N_{n_{r}\kappa}+\kappa')-C_{k}^{(2)}(n_{r}-k-3) \right],
\label{eq:6}
\end{equation}
\begin{equation}
\mathcal{Y}(p)=\frac{(-)^{p}}{p!(n_{r}-p)!} \frac{\Gamma(\gamma_{\kappa}+\gamma_{\kappa'}+p+2)}{\Gamma(p+2\gamma_{\kappa}+1)}\left[(n_{r}-p)(\kappa+\kappa')+2(N_{n_{r}\kappa}-\kappa) \right],
\label{eq:7}
\end{equation}
with
\begin{equation}
C_{k}^{(1)}=(n_{r}-k)+\frac{n_{r}+\gamma_{\kappa}}{N_{n_{r}\kappa}}\left(\kappa-N_{n_{r}\kappa}\right), \qquad  \qquad C_{k}^{(2)}=\frac{n_{r}+\gamma_{\kappa}}{N_{n_{r}\kappa}}(n_{r}-k)+\left(\kappa-N_{n_{r}\kappa}\right),
\label{eq:8}
\end{equation}
\begin{equation}
N_{n_{r}\kappa}=\sqrt{n_{r}^2+2n_{r}\gamma_{\kappa}+\kappa^2}
\label{eq:9}
\end{equation}
and
\begin{equation}
\gamma_{\kappa}=\sqrt{\kappa^2-(\alpha Z)^2}
\label{eq:10}.
\end{equation}
In Ref.\ \cite{Stef16} we have shown that the expression from Eq.\ (\ref{eq:2}) remains valid for an \emph{arbitrary} discrete energy eigenstate of the atom. Nevertheless, we realize that its quite complexity form might be perceived by someone as inconvenient for calculations. Such our thoughts, as well as the lack of any tabulated data in the aforementioned article, were the main reasons for which we decided to perform numerical calculations of that quantity. In the next Section we will discuss briefly our results.

\section{Discussion of results}

The computational results obtained with the aid of the formula in Eq.\ (\ref{eq:2}) are presented here in the form of 30 tables comprising the values of the $\chi_{\textrm{M}1 \to \textrm{E}2}$  for the relativistic hydrogenlike atoms (with the nuclear charge number from the range $1 \leqslant Z \leqslant 137$) being in the ground state 1s$_{1/2}$, or in any state belonging to the first three sets of excited states, i.e.: 2s$_{1/2}$, 2p$_{1/2}$, 2p$_{3/2}$, 3s$_{1/2}$, 3p$_{1/2}$, 3p$_{3/2}$, 3d$_{3/2}$, 3d$_{5/2}$, 4s$_{1/2}$, 4p$_{1/2}$, 4p$_{3/2}$, 4d$_{3/2}$, 4d$_{5/2}$, 4f$_{5/2}$ and 4f$_{7/2}$, having regarded all possible values of the magnetic quantum number $\mu$. 

All tabulated data were obtained with the use of the current value $\alpha^{-1}=137.035 \: 999 \: 139$ of the inverse of the fine-structure constant recommended by the Committee on Data for Science and Technology (CODATA 2014) \cite{Mohr14}.

Values for the susceptibility in question for the atomic ground state (Table \ref{tab:1s_1-2}) fully coincide with those obtained by Szmytkowski and {\L}ukasik \cite{Szmy16b}. Analysis of all numbers from Tables \ref{tab:1s_1-2}--\ref{tab:4f_7-2_mu_7-2} allows one to conclude that the sign of $\chi_{\textrm{M}1 \to \textrm{E}2}$ is related to the sign of the angular-plus-parity symmetry quantum number $\kappa$, i.e.: generally, if $\kappa>0$ then $\chi_{\textrm{M}1 \to \textrm{E}2}$ takes the positive values, and if $\kappa<0$ then  $\chi_{\textrm{M}1 \to \textrm{E}2}$ is negative. Exceptions to this rule are states 1s$_{1/2}$, 2s$_{1/2}$ and those of the type $n$p$_{3/2}$, with $|\mu|=1/2$. For these latter there is the critical value of nuclear charge number $Z$, $Z_c$, such that for $Z<Z_c$ the $\chi_{\textrm{M}1 \to \textrm{E}2}$ susceptibility is negative, and for $Z \geqslant Z_c$ it takes the positive values. For the state 2p$_{3/2}$ it equals  $Z_c=123$, and increases for the higher excited states, i.e. for the 3p$_{3/2}$ state there is $Z_c=127$, and for the state 4p$_{3/2}$ we have $Z_c=129$. Furthermore, for states with orbital angular momentum $l=2$ (the d-type), the values of $\chi_{\textrm{M}1 \to \textrm{E}2}$ for every ions with $1 \leqslant Z \leqslant 137$, calculated  with $|\mu|=3/2$, are approximately two times greater than appropriate numbers obtained with the use of $|\mu|=1/2$. Moreover, the susceptibility under study for d-type states with total angular momentum $j=3/2$ takes nearly the same values as those with $j=5/2$, but with the opposite sign (this is caused by opposite directed spins). In turn, in determination of $\chi_{\textrm{M}1 \to \textrm{E}2}$ for the f-type states, there is no matter, which magnetic quantum number is taken into account: differences between values obtained with $|\mu|=3/2$ and those with $|\mu|=5/2$ are of the order of $10^{-8}$ in the units of $a_0^{4}$.



\ack
I would like to thank Professor A.\ Rutkowski for his strong suggestion to publish present results. Useful discussions with Professor R.\ Szmytkowski are also acknowledged.



\clearpage

\newpage

\TableExplanation

In all the tables we have used the following notation:

\bigskip
\renewcommand{\arraystretch}{1.0}

\begin{tabular}{ll}
$\chi_{\textrm{M}1 \to \textrm{E}2}$ 				& The magnetic-dipole-to-electric-quadrupole atomic cross-susceptibility \\
$Z$ 					& The nuclear charge number \\
$\alpha$ 			& The fine structure constant \\
$a_0$					& The Bohr radius \\
$a(b)$				& This means $a \times 10^b$ \\
$\mu$         & The magnetic quantum number; $\mu=-j, \ldots, j$ \\*[1.5ex]
\multicolumn{2}{l}{The atomic states are referred to as $n\textrm{x}_{j}$, due to the following guidelines:} \\*[1ex]
$n$  					& The principal quantum number \\
$n=n_{r}+|\kappa|$ & Note that for the radial quantum number $n_{r}=0$, $\kappa$ is the negative integer! \\
$\kappa$ 			& The Dirac quantum number; $\kappa=(l-j)(2j+1)$\\
$j$						& The total angular momentum quantum number; $j=|\kappa|-\frac{1}{2}=l \pm \frac{1}{2}$ \\
x $\equiv$ s	& States with the orbital angular momentum quantum number $l=0$ \\
x $\equiv$ p	& States with the orbital angular momentum quantum number  $l=1$ \\
x $\equiv$ d	& States with the orbital angular momentum quantum number  $l=2$ \\
x $\equiv$ f	& States with the orbital angular momentum quantum number  $l=3$ \\
\end{tabular}

\datatables 



\setlength{\LTleft}{0pt}
\setlength{\LTright}{0pt} 


\setlength{\tabcolsep}{0.5\tabcolsep}

\renewcommand{\arraystretch}{1.0}

\footnotesize 

\begin{longtable}{@{\extracolsep\fill}rlrlrl@{}}
\caption{The magnetic-dipole-to-electric-quadrupole cross-susceptibilities $\chi_{\textrm{M}1 \to \textrm{E}2}$ for the ground state (1s$_{1/2}$) of Dirac one-electron atoms, computed from the analytical formula in Eq.\ (\ref{eq:2}). Notation $a(b)$ denotes $a \times 10^{b}$. The value of the inverse of the fine-structure constant used in the calculations is $\alpha^{-1}=137.035 \: 999 \: 139$ (from CODATA 2014).}
\label{tab:1s_1-2}
$Z$ & \multicolumn{1}{c}{$\chi_{\textrm{M}1 \to \textrm{E}2}[a_0^4]$} & $Z$ & \multicolumn{1}{c}{$\chi_{\textrm{M}1 \to \textrm{E}2}[a_0^4]$} & $Z$ & \multicolumn{1}{c}{$\chi_{\textrm{M}1 \to \textrm{E}2}[a_0^4]$} \\
\hline\\
\endfirsthead
\caption[]{(continued)}
$Z$ & \multicolumn{1}{c}{$\chi_{\textrm{M}1 \to \textrm{E}2}[a_0^4]$} & $Z$ & \multicolumn{1}{c}{$\chi_{\textrm{M}1 \to \textrm{E}2}[a_0^4]$} & $Z$ & \multicolumn{1}{c}{$\chi_{\textrm{M}1 \to \textrm{E}2}[a_0^4]$} \\
\hline\\
\endhead
1    &   7.447 801 428 672($-$8)    &   47   &   3.120 711 353 217($-$11)   &   93    &   6.002 694 384 905($-$12)                              \\

2    &   1.861 763 994 268($-$8)    &   48   &   2.981 572 919 250($-$11)   &   94    &   5.816 695 855 231($-$12)                              \\

3    &   8.273 126 128 262($-$9)    &   49   &   2.850 855 824 355($-$11)   &   95    &   5.636 311 517 841($-$12)                              \\

4    &   4.652 546 237 671($-$9)    &   50   &   2.727 892 843 009($-$11)   &   96    &   5.461 299 261 670($-$12)                              \\

5    &   2.976 734 913 049($-$9)    &   51   &   2.612 081 508 701($-$11)   &   97    &   5.291 428 810 951($-$12)                              \\

6    &   2.066 417 585 548($-$9)    &   52   &   2.502 876 714 228($-$11)   &   98    &   5.126 480 952 929($-$12)                              \\

7    &   1.517 524 880 849($-$9)    &   53   &   2.399 784 279 927($-$11)   &   99    &   4.966 246 813 946($-$12)                              \\

8    &   1.161 272 334 573($-$9)    &   54   &   2.302 355 347 783($-$11)   &   100   &   4.810 527 179 193($-$12)                              \\
\hline
\end{longtable}

\newpage

\footnotesize
\begin{longtable}{@{\extracolsep\fill}rlrlrl@{}}
\caption{The magnetic-dipole-to-electric-quadrupole cross-susceptibilities $\chi_{\textrm{M}1 \to \textrm{E}2}$ for Dirac one-electron atoms in the excited state 2s$_{1/2}$, computed from the analytical formula in Eq.\ (\ref{eq:2}). Notation $a(b)$ denotes $a \times 10^{b}$. The value of the inverse of the fine-structure constant used in the calculations is $\alpha^{-1}=137.035 \: 999 \: 139$ (from CODATA 2014).}
\label{tab:2s_1-2}
$Z$ & \multicolumn{1}{c}{$\chi_{\textrm{M}1 \to \textrm{E}2}[a_0^4]$} & $Z$ & \multicolumn{1}{c}{$\chi_{\textrm{M}1 \to \textrm{E}2}[a_0^4]$} & $Z$ & \multicolumn{1}{c}{$\chi_{\textrm{M}1 \to \textrm{E}2}[a_0^4]$} \\
\hline\\
\endfirsthead
\caption[]{(continued)}
$Z$ & \multicolumn{1}{c}{$\chi_{\textrm{M}1 \to \textrm{E}2}[a_0^4]$} & $Z$ & \multicolumn{1}{c}{$\chi_{\textrm{M}1 \to \textrm{E}2}[a_0^4]$} & $Z$ & \multicolumn{1}{c}{$\chi_{\textrm{M}1 \to \textrm{E}2}[a_0^4]$} \\
\hline\\
\endhead
1    &   2.202 146 300 425($-$7)    &   47   &   1.099 649 053 421($-$10)   &   93    &   3.212 961 137 406($-$11)                              \\

2    &   5.506 221 751 591($-$8)    &   48   &   1.058 048 926 220($-$10)   &   94    &   3.147 644 917 689($-$11)                              \\

3    &   2.447 843 517 190($-$8)    &   49   &   1.018 928 194 936($-$10)   &   95    &   3.083 825 182 331($-$11)                              \\

4    &   1.377 410 869 756($-$8)    &   50   &   9.820 895 472 425($-$11)   &   96    &   3.021 427 034 310($-$11)                              \\

5    &   8.819 531 817 233($-$9)    &   51   &   9.473 548 152 849($-$11)   &   97    &   2.960 378 869 994($-$11)                              \\

6    &   6.128 153 631 425($-$9)    &   52   &   9.145 627 878 784($-$11)   &   98    &   2.900 612 132 845($-$11)                              \\

7    &   4.505 334 250 484($-$9)    &   53   &   8.835 673 088 781($-$11)   &   99    &   2.842 061 079 345($-$11)                              \\

8    &   3.452 058 735 804($-$9)    &   54   &   8.542 356 197 163($-$11)   &   100   &   2.784 662 555 448($-$11)                              \\
\hline
\end{longtable}

\newpage

\footnotesize
\begin{longtable}{@{\extracolsep\fill}rlrlrl@{}}
\caption{The magnetic-dipole-to-electric-quadrupole cross-susceptibilities $\chi_{\textrm{M}1 \to \textrm{E}2}$ for Dirac one-electron atoms in the excited state 2p$_{1/2}$, computed from the analytical formula in Eq.\ (\ref{eq:2}). Notation $a(b)$ denotes $a \times 10^{b}$. The value of the inverse of the fine-structure constant used in the calculations is $\alpha^{-1}=137.035 \: 999 \: 139$ (from CODATA 2014).}
\label{tab:2p_1-2}
$Z$ & \multicolumn{1}{c}{$\chi_{\textrm{M}1 \to \textrm{E}2}[a_0^4]$} & $Z$ & \multicolumn{1}{c}{$\chi_{\textrm{M}1 \to \textrm{E}2}[a_0^4]$} & $Z$ & \multicolumn{1}{c}{$\chi_{\textrm{M}1 \to \textrm{E}2}[a_0^4]$} \\
\hline\\
\endfirsthead
\caption[]{(continued)}
$Z$ & \multicolumn{1}{c}{$\chi_{\textrm{M}1 \to \textrm{E}2}[a_0^4]$} & $Z$ & \multicolumn{1}{c}{$\chi_{\textrm{M}1 \to \textrm{E}2}[a_0^4]$} & $Z$ & \multicolumn{1}{c}{$\chi_{\textrm{M}1 \to \textrm{E}2}[a_0^4]$} \\
\hline\\
\endhead
1    &   1.753 947 534 759($+$4)    &   47   &   1.398 607 367 271($-$6)   &   93    &   1.307 597 021 456($-$8)                               \\

2    &   2.740 012 156 862($+$2)    &   48   &   1.224 120 137 260($-$6)   &   94    &   1.200 781 095 853($-$8)                               \\

3    &   2.404 720 920 131($+$1)    &   49   &   1.073 997 620 958($-$6)   &   95    &   1.102 827 220 706($-$8)                               \\

4    &   4.277 951 385 360($+$0)    &   50   &   9.444 668 526 493($-$7)   &   96    &   1.012 953 427 235($-$8)                               \\

5    &   1.120 787 103 116($+$0)    &   51   &   8.323 960 185 841($-$7)   &   97    &   9.304 522 704 072($-$9)                               \\

6    &   3.750 824 136 872($-$1)   &   52   &   7.351 760 467 397($-$7)   &   98    &   8.546 831 473 414($-$9)                               \\

7    &   1.486 211 494 659($-$1)   &   53   &   6.506 257 637 804($-$7)   &   99    &   7.850 654 691 477($-$9)                               \\

8    &   6.663 571 600 499($-$2)   &   54   &   5.769 156 006 640($-$7)   &   100   &   7.210 725 844 452($-$9)                               \\
\hline
\end{longtable}

\newpage

\footnotesize
\begin{longtable}{@{\extracolsep\fill}rlrlrl@{}}
\caption{The magnetic-dipole-to-electric-quadrupole cross-susceptibilities $\chi_{\textrm{M}1 \to \textrm{E}2}$ for Dirac one-electron atoms in the excited state 2p$_{3/2}$ ($|\mu|=1/2$), computed from the analytical formula in Eq.\ (\ref{eq:2}). Notation $a(b)$ denotes $a \times 10^{b}$. The value of the inverse of the fine-structure constant used in the calculations is $\alpha^{-1}=137.035 \: 999 \: 139$ (from CODATA 2014).}
\label{tab:2p_3-2_mu_1-2}
$Z$ & \multicolumn{1}{c}{$\chi_{\textrm{M}1 \to \textrm{E}2}[a_0^4]$} & $Z$ & \multicolumn{1}{c}{$\chi_{\textrm{M}1 \to \textrm{E}2}[a_0^4]$} & $Z$ & \multicolumn{1}{c}{$\chi_{\textrm{M}1 \to \textrm{E}2}[a_0^4]$} \\
\hline\\
\endfirsthead
\caption[]{(continued)}
$Z$ & \multicolumn{1}{c}{$\chi_{\textrm{M}1 \to \textrm{E}2}[a_0^4]$} & $Z$ & \multicolumn{1}{c}{$\chi_{\textrm{M}1 \to \textrm{E}2}[a_0^4]$} & $Z$ & \multicolumn{1}{c}{$\chi_{\textrm{M}1 \to \textrm{E}2}[a_0^4]$} \\
\hline\\
\endhead
1    &   $-$1.753 947 534 455($+$4)    &   47   &   $-$1.396 758 216 281($-$6)   &   93    &   $-$1.219 044 245 690($-$8)                            \\

2    &   $-$2.740 012 149 266($+$2)    &   48   &   $-$1.222 327 073 405($-$6)   &   94    &   $-$1.112 709 022 228($-$8)                            \\

3    &   $-$2.404 720 886 345($+$1)    &   49   &   $-$1.072 257 190 831($-$6)   &   95    &   $-$1.015 202 035 545($-$8)                            \\

4    &   $-$4.277 951 195 111($+$0)    &   50   &   $-$9.427 758 743 717($-$7)   &   96    &   $-$9.257 416 926 879($-$9)                            \\

5    &   $-$1.120 786 981 191($+$0)    &   51   &   $-$8.307 515 553 759($-$7)   &   97    &   $-$8.436 208 274 524($-$9)                            \\

6    &   $-$3.750 823 288 752($-$1)   &   52   &   $-$7.335 753 835 131($-$7)   &   98    &   $-$7.681 990 168 215($-$9)                            \\

7    &   $-$1.486 210 870 323($-$1)   &   53   &   $-$6.490 663 863 521($-$7)   &   99    &   $-$6.988 957 499 715($-$9)                            \\

8    &   $-$6.663 566 809 580($-$2)   &   54   &   $-$5.753 951 772 515($-$7)   &   100   &   $-$6.351 843 479 901($-$9)                            \\
\hline
\end{longtable}

\newpage

\footnotesize
\begin{longtable}{@{\extracolsep\fill}rlrlrl@{}}
\caption{The magnetic-dipole-to-electric-quadrupole cross-susceptibilities $\chi_{\textrm{M}1 \to \textrm{E}2}$ for Dirac one-electron atoms in the excited state 2p$_{3/2}$ ($|\mu|=3/2$), computed from the analytical formula in Eq.\ (\ref{eq:2}). Notation $a(b)$ denotes $a \times 10^{b}$. The value of the inverse of the fine-structure constant used in the calculations is $\alpha^{-1}=137.035 \: 999 \: 139$ (from CODATA 2014).}
\label{tab:2p_3-2_mu_3-2}
$Z$ & \multicolumn{1}{c}{$\chi_{\textrm{M}1 \to \textrm{E}2}[a_0^4]$} & $Z$ & \multicolumn{1}{c}{$\chi_{\textrm{M}1 \to \textrm{E}2}[a_0^4]$} & $Z$ & \multicolumn{1}{c}{$\chi_{\textrm{M}1 \to \textrm{E}2}[a_0^4]$} \\
\hline\\
\endfirsthead
\caption[]{(continued)}
$Z$ & \multicolumn{1}{c}{$\chi_{\textrm{M}1 \to \textrm{E}2}[a_0^4]$} & $Z$ & \multicolumn{1}{c}{$\chi_{\textrm{M}1 \to \textrm{E}2}[a_0^4]$} & $Z$ & \multicolumn{1}{c}{$\chi_{\textrm{M}1 \to \textrm{E}2}[a_0^4]$} \\
\hline\\
\endhead
1    &   $-$3.253 055 437 887($-$6)   &   47   &   $-$1.441 867 490 876($-$9)    &   93    &   $-$3.451 069 326 858($-$10)   \\

2    &   $-$8.132 408 280 706($-$7)   &   48   &   $-$1.381 142 920 224($-$9)    &   94    &   $-$3.371 398 250 998($-$10)   \\

3    &   $-$3.614 233 075 851($-$7)   &   49   &   $-$1.324 097 838 036($-$9)    &   95    &   $-$3.294 226 672 660($-$10)   \\

4    &   $-$2.032 871 752 390($-$7)   &   50   &   $-$1.270 440 843 186($-$9)    &   96    &   $-$3.219 450 944 023($-$10)   \\

5    &   $-$1.300 927 366 455($-$7)   &   51   &   $-$1.219 908 820 279($-$9)    &   97    &   $-$3.146 972 731 748($-$10)   \\

6    &   $-$9.033 279 448 830($-$8)   &   52   &   $-$1.172 263 707 719($-$9)    &   98    &   $-$3.076 698 693 259($-$10)   \\

7    &   $-$6.635 880 317 790($-$8)   &   53   &   $-$1.127 289 688 545($-$9)    &   99    &   $-$3.008 540 175 802($-$10)   \\

8    &   $-$5.079 876 052 049($-$8)   &   54   &   $-$1.084 790 741 988($-$9)    &   100   &   $-$2.942 412 936 458($-$10)   \\
\hline
\end{longtable}

\newpage

\footnotesize
\begin{longtable}{@{\extracolsep\fill}rlrlrl@{}}
\caption{The magnetic-dipole-to-electric-quadrupole cross-susceptibilities $\chi_{\textrm{M}1 \to \textrm{E}2}$ for Dirac one-electron atoms in the excited state 3s$_{1/2}$, computed from the analytical formula in Eq.\ (\ref{eq:2}). Notation $a(b)$ denotes $a \times 10^{b}$. The value of the inverse of the fine-structure constant used in the calculations is $\alpha^{-1}=137.035 \: 999 \: 139$ (from CODATA 2014).}
\label{tab:3s_1-2}
$Z$ & \multicolumn{1}{c}{$\chi_{\textrm{M}1 \to \textrm{E}2}[a_0^4]$} & $Z$ & \multicolumn{1}{c}{$\chi_{\textrm{M}1 \to \textrm{E}2}[a_0^4]$} & $Z$ & \multicolumn{1}{c}{$\chi_{\textrm{M}1 \to \textrm{E}2}[a_0^4]$} \\
\hline\\
\endfirsthead
\caption[]{(continued)}
$Z$ & \multicolumn{1}{c}{$\chi_{\textrm{M}1 \to \textrm{E}2}[a_0^4]$} & $Z$ & \multicolumn{1}{c}{$\chi_{\textrm{M}1 \to \textrm{E}2}[a_0^4]$} & $Z$ & \multicolumn{1}{c}{$\chi_{\textrm{M}1 \to \textrm{E}2}[a_0^4]$} \\
\hline\\
\endhead
1    &   $-$1.313 491 400 853($-$1)   &   47   &   $-$2.526 248 540 764($-$8)   &   93    &   $-$1.194 912 283 120($-$9)                              \\

2    &   $-$8.208 719 952 903($-$3)   &   48   &   $-$2.314 848 128 289($-$8)   &   94    &   $-$1.129 620 134 730($-$9)                              \\

3    &   $-$1.621 277 483 253($-$3)   &   49   &   $-$2.124 597 516 202($-$8)   &   95    &   $-$1.067 894 183 379($-$9)                              \\

4    &   $-$5.128 945 243 042($-$4)   &   50   &   $-$1.953 020 927 338($-$8)   &   96    &   $-$1.009 510 127 908($-$9)                              \\

5    &   $-$2.100 353 086 575($-$4)   &   51   &   $-$1.797 973 469 976($-$8)   &   97    &   $-$9.542 597 847 388($-$10)                             \\

6    &   $-$1.012 628 569 669($-$4)   &   52   &   $-$1.657 591 603 708($-$8)   &   98    &   $-$9.019 497 797 699($-$10)                             \\

7    &   $-$5.464 171 378 232($-$5)   &   53   &   $-$1.530 251 788 631($-$8)   &   99    &   $-$8.524 003 576 936($-$10)                             \\

8    &   $-$3.201 814 253 364($-$5)   &   54   &   $-$1.414 535 843 462($-$8)   &   100   &   $-$8.054 442 970 884($-$10)                             \\
\hline
\end{longtable}

\newpage

\footnotesize
\begin{longtable}{@{\extracolsep\fill}rlrlrc@{}}
\caption{The magnetic-dipole-to-electric-quadrupole cross-susceptibilities $\chi_{\textrm{M}1 \to \textrm{E}2}$ for Dirac one-electron atoms in the excited state 3p$_{1/2}$, computed from the analytical formula in Eq.\ (\ref{eq:2}). Notation $a(b)$ denotes $a \times 10^{b}$. The value of the inverse of the fine-structure constant used in the calculations is $\alpha^{-1}=137.035 \: 999 \: 139$ (from CODATA 2014).}
\label{tab:3p_1-2}
$Z$ & \multicolumn{1}{c}{$\chi_{\textrm{M}1 \to \textrm{E}2}[a_0^4]$} & $Z$ & \multicolumn{1}{c}{$\chi_{\textrm{M}1 \to \textrm{E}2}[a_0^4]$} & $Z$ & \multicolumn{1}{c}{$\chi_{\textrm{M}1 \to \textrm{E}2}[a_0^4]$} \\
\hline\\
\endfirsthead
\caption[]{(continued)}
$Z$ & \multicolumn{1}{c}{$\chi_{\textrm{M}1 \to \textrm{E}2}[a_0^4]$} & $Z$ & \multicolumn{1}{c}{$\chi_{\textrm{M}1 \to \textrm{E}2}[a_0^4]$} & $Z$ & \multicolumn{1}{c}{$\chi_{\textrm{M}1 \to \textrm{E}2}[a_0^4]$} \\
\hline\\
\endhead
1    &   3.551 771 997 529($+$5)    &   47   &   2.884 781 726 595($-$5)   &   93    &   2.873 432 400 052($-$7)                              \\

2    &   5.548 701 094 565($+$3)    &   48   &   2.526 978 494 602($-$5)   &   94    &   2.644 043 937 472($-$7)                              \\

3    &   4.869 908 391 007($+$2)    &   49   &   2.218 963 742 422($-$5)   &   95    &   2.433 333 725 049($-$7)                              \\

4    &   8.663 954 014 757($+$1)    &   50   &   1.953 044 363 637($-$5)   &   96    &   2.239 671 340 508($-$7)                              \\

5    &   2.270 045 290 951($+$1)    &   51   &   1.722 832 661 853($-$5)   &   97    &   2.061 580 664 162($-$7)                              \\

6    &   7.597 594 857 192($+$0)    &   52   &   1.523 005 509 195($-$5)   &   98    &   1.897 724 046 194($-$7)                              \\

7    &   3.010 752 213 461($+$0)    &   53   &   1.349 111 403 750($-$5)   &   99    &   1.746 888 227 061($-$7)                              \\

8    &   1.350 060 981 160($+$0)    &   54   &   1.197 415 231 798($-$5)   &   100   &   1.607 971 802 434($-$7)                              \\
\hline
\end{longtable}

\newpage

\footnotesize
\begin{longtable}{@{\extracolsep\fill}rlrlrl@{}}
\caption{The magnetic-dipole-to-electric-quadrupole cross-susceptibilities $\chi_{\textrm{M}1 \to \textrm{E}2}$ for Dirac one-electron atoms in the excited state 3p$_{3/2}$ ($|\mu|=1/2$), computed from the analytical formula in Eq.\ (\ref{eq:2}). Notation $a(b)$ denotes $a \times 10^{b}$. The value of the inverse of the fine-structure constant used in the calculations is $\alpha^{-1}=137.035 \: 999 \: 139$ (from CODATA 2014).}
\label{tab:3p_3-2_mu_1-2}
$Z$ & \multicolumn{1}{c}{$\chi_{\textrm{M}1 \to \textrm{E}2}[a_0^4]$} & $Z$ & \multicolumn{1}{c}{$\chi_{\textrm{M}1 \to \textrm{E}2}[a_0^4]$} & $Z$ & \multicolumn{1}{c}{$\chi_{\textrm{M}1 \to \textrm{E}2}[a_0^4]$} \\
\hline\\
\endfirsthead
\caption[]{(continued)}
$Z$ & \multicolumn{1}{c}{$\chi_{\textrm{M}1 \to \textrm{E}2}[a_0^4]$} & $Z$ & \multicolumn{1}{c}{$\chi_{\textrm{M}1 \to \textrm{E}2}[a_0^4]$} & $Z$ & \multicolumn{1}{c}{$\chi_{\textrm{M}1 \to \textrm{E}2}[a_0^4]$} \\
\hline\\
\endhead
1    &   $-$3.551 771 997 357($+$5)    &   47   &   $-$2.883 319 793 463($-$5)   &   93    &   $-$2.772 313 534 646($-$7)                            \\

2    &   $-$5.548 701 090 259($+$3)    &   48   &   $-$2.525 547 321 643($-$5)   &   94    &   $-$2.542 802 270 856($-$7)                            \\

3    &   $-$4.869 908 371 831($+$2)    &   49   &   $-$2.217 561 343 867($-$5)   &   95    &   $-$2.331 935 582 426($-$7)                            \\

4    &   $-$8.663 953 906 605($+$1)    &   50   &   $-$1.951 668 906 857($-$5)   &   96    &   $-$2.138 082 469 294($-$7)                            \\

5    &   $-$2.270 045 221 498($+$1)    &   51   &   $-$1.721 482 452 365($-$5)   &   97    &   $-$1.959 766 115 935($-$7)                            \\

6    &   $-$7.597 594 372 875($+$0)    &   52   &   $-$1.521 678 977 369($-$5)   &   98    &   $-$1.795 648 050 905($-$7)                            \\

7    &   $-$3.010 751 855 896($+$0)    &   53   &   $-$1.347 807 093 003($-$5)   &   99    &   $-$1.644 514 058 120($-$7)                            \\

8    &   $-$1.350 060 705 861($+$0)    &   54   &   $-$1.196 131 788 085($-$5)   &   100   &   $-$1.505 261 631 112($-$7)                            \\
\hline
\end{longtable}

\newpage

\footnotesize
\begin{longtable}{@{\extracolsep\fill}rlrlrl@{}}
\caption{The magnetic-dipole-to-electric-quadrupole cross-susceptibilities $\chi_{\textrm{M}1 \to \textrm{E}2}$ for Dirac one-electron atoms in the excited state 3p$_{3/2}$ ($|\mu|=3/2$), computed from the analytical formula in Eq.\ (\ref{eq:2}). Notation $a(b)$ denotes $a \times 10^{b}$. The value of the inverse of the fine-structure constant used in the calculations is $\alpha^{-1}=137.035 \: 999 \: 139$ (from CODATA 2014).}
\label{tab:3p_3-2_mu_3-2}
$Z$ & \multicolumn{1}{c}{$\chi_{\textrm{M}1 \to \textrm{E}2}[a_0^4]$} & $Z$ & \multicolumn{1}{c}{$\chi_{\textrm{M}1 \to \textrm{E}2}[a_0^4]$} & $Z$ & \multicolumn{1}{c}{$\chi_{\textrm{M}1 \to \textrm{E}2}[a_0^4]$} \\
\hline\\
\endfirsthead
\caption[]{(continued)}
$Z$ & \multicolumn{1}{c}{$\chi_{\textrm{M}1 \to \textrm{E}2}[a_0^4]$} & $Z$ & \multicolumn{1}{c}{$\chi_{\textrm{M}1 \to \textrm{E}2}[a_0^4]$} & $Z$ & \multicolumn{1}{c}{$\chi_{\textrm{M}1 \to \textrm{E}2}[a_0^4]$} \\
\hline\\
\endhead
1    &   $-$2.155 838 547 139($-$5)   &   47   &   $-$9.510 511 813 286($-$9)   &   93    &   $-$2.245 052 801 466($-$9)    \\

2    &   $-$5.389 409 299 981($-$6)   &   48   &   $-$9.108 124 397 344($-$9)   &   94    &   $-$2.192 339 914 006($-$9)    \\

3    &   $-$2.395 154 454 398($-$6)   &   49   &   $-$8.730 122 353 534($-$9)   &   95    &   $-$2.141 284 484 492($-$9)    \\

4    &   $-$1.347 165 259 595($-$6)   &   50   &   $-$8.374 574 529 079($-$9)   &   96    &   $-$2.091 817 827 216($-$9)    \\

5    &   $-$8.620 959 764 062($-$7)   &   51   &   $-$8.039 737 224 533($-$9)   &   97    &   $-$2.043 874 778 500($-$9)    \\

6    &   $-$5.986 015 523 099($-$7)   &   52   &   $-$7.724 032 775 347($-$9)   &   98    &   $-$1.997 393 482 169($-$9)    \\

7    &   $-$4.397 227 265 605($-$7)   &   53   &   $-$7.426 030 935 159($-$9)   &   99    &   $-$1.952 315 190 105($-$9)    \\

8    &   $-$3.366 042 593 635($-$7)   &   54   &   $-$7.144 432 649 641($-$9)   &   100   &   $-$1.908 584 076 706($-$9)    \\
\hline
\end{longtable}

\newpage

\footnotesize
\begin{longtable}{@{\extracolsep\fill}rlrlrl@{}}
\caption{The magnetic-dipole-to-electric-quadrupole cross-susceptibilities $\chi_{\textrm{M}1 \to \textrm{E}2}$ for Dirac one-electron atoms in the excited state 3d$_{3/2}$ ($|\mu|=1/2$), computed from the analytical formula in Eq.\ (\ref{eq:2}). Notation $a(b)$ denotes $a \times 10^{b}$. The value of the inverse of the fine-structure constant used in the calculations is $\alpha^{-1}=137.035 \: 999 \: 139$ (from CODATA 2014).}
\label{tab:3d_3-2_mu_1-2}
$Z$ & \multicolumn{1}{c}{$\chi_{\textrm{M}1 \to \textrm{E}2}[a_0^4]$} & $Z$ & \multicolumn{1}{c}{$\chi_{\textrm{M}1 \to \textrm{E}2}[a_0^4]$} & $Z$ & \multicolumn{1}{c}{$\chi_{\textrm{M}1 \to \textrm{E}2}[a_0^4]$} \\
\hline\\
\endfirsthead
\caption[]{(continued)}
$Z$ & \multicolumn{1}{c}{$\chi_{\textrm{M}1 \to \textrm{E}2}[a_0^4]$} & $Z$ & \multicolumn{1}{c}{$\chi_{\textrm{M}1 \to \textrm{E}2}[a_0^4]$} & $Z$ & \multicolumn{1}{c}{$\chi_{\textrm{M}1 \to \textrm{E}2}[a_0^4]$} \\
\hline\\
\endhead
1    &   1.918 026 660 215($+$5)    &   47   &   1.700 802 430 874($-$5)   &   93    &   2.468 430 348 548($-$7)                            \\

2    &   2.996 734 726 747($+$3)    &   48   &   1.496 029 028 709($-$5)   &   94    &   2.305 376 415 585($-$7)                            \\

3    &   2.630 612 036 988($+$2)    &   49   &   1.319 284 840 518($-$5)   &   95    &   2.154 437 322 608($-$7)                            \\

4    &   4.681 261 040 683($+$1)    &   50   &   1.166 282 488 906($-$5)   &   96    &   2.014 608 000 663($-$7)                            \\

5    &   1.226 940 982 099($+$1)    &   51   &   1.033 457 249 038($-$5)   &   97    &   1.884 975 564 618($-$7)                            \\

6    &   4.108 085 743 031($+$0)    &   52   &   9.178 343 452 506($-$6)   &   98    &   1.764 710 052 313($-$7)                            \\

7    &   1.628 712 652 878($+$0)    &   53   &   8.169 225 501 298($-$6)   &   99    &   1.653 056 174 167($-$7)                            \\

8    &   7.307 366 604 495($-$1)   &   54   &   7.286 285 025 304($-$6)   &   100   &   1.549 325 954 352($-$7)                            \\
\hline
\end{longtable}

\newpage

\footnotesize
\begin{longtable}{@{\extracolsep\fill}rlrlrl@{}}
\caption{The magnetic-dipole-to-electric-quadrupole cross-susceptibilities $\chi_{\textrm{M}1 \to \textrm{E}2}$ for Dirac one-electron atoms in the excited state 3d$_{3/2}$ ($|\mu|=3/2$), computed from the analytical formula in Eq.\ (\ref{eq:2}). Notation $a(b)$ denotes $a \times 10^{b}$. The value of the inverse of the fine-structure constant used in the calculations is $\alpha^{-1}=137.035 \: 999 \: 139$ (from CODATA 2014).}
\label{tab:3d_3-2_mu_3-2}
$Z$ & \multicolumn{1}{c}{$\chi_{\textrm{M}1 \to \textrm{E}2}[a_0^4]$} & $Z$ & \multicolumn{1}{c}{$\chi_{\textrm{M}1 \to \textrm{E}2}[a_0^4]$} & $Z$ & \multicolumn{1}{c}{$\chi_{\textrm{M}1 \to \textrm{E}2}[a_0^4]$} \\
\hline\\
\endfirsthead
\caption[]{(continued)}
$Z$ & \multicolumn{1}{c}{$\chi_{\textrm{M}1 \to \textrm{E}2}[a_0^4]$} & $Z$ & \multicolumn{1}{c}{$\chi_{\textrm{M}1 \to \textrm{E}2}[a_0^4]$} & $Z$ & \multicolumn{1}{c}{$\chi_{\textrm{M}1 \to \textrm{E}2}[a_0^4]$} \\
\hline\\
\endhead
1    &   3.836 050 693 130($+$5)    &   47   &   3.395 129 533 128($-$5)   &   93    &   4.878 422 654 270($-$7)   \\

2    &   5.993 453 028 098($+$3)    &   48   &   2.986 064 993 086($-$5)   &   94    &   4.554 416 069 391($-$7)   \\

3    &   5.261 191 613 066($+$2)    &   49   &   2.633 012 983 039($-$5)   &   95    &   4.254 543 818 061($-$7)   \\

4    &   9.362 419 303 567($+$1)    &   50   &   2.327 403 984 856($-$5)   &   96    &   3.976 801 281 494($-$7)   \\

5    &   2.453 839 829 849($+$1)    &   51   &   2.062 113 096 082($-$5)   &   97    &   3.719 367 835 655($-$7)   \\

6    &   8.215 968 076 462($+$0)    &   52   &   1.831 194 725 393($-$5)   &   98    &   3.480 588 359 046($-$7)   \\

7    &   3.257 315 372 988($+$0)    &   53   &   1.629 669 876 199($-$5)   &   99    &   3.258 956 758 836($-$7)   \\

8    &   1.461 408 787 502($+$0)    &   54   &   1.453 354 854 369($-$5)   &   100   &   3.053 101 277 849($-$7)   \\
\hline
\end{longtable}

\newpage

\footnotesize
\begin{longtable}{@{\extracolsep\fill}rlrlrl@{}}
\caption{The magnetic-dipole-to-electric-quadrupole cross-susceptibilities $\chi_{\textrm{M}1 \to \textrm{E}2}$ for Dirac one-electron atoms in the excited state 3d$_{5/2}$ ($|\mu|=1/2$), computed from the analytical formula in Eq.\ (\ref{eq:2}). Notation $a(b)$ denotes $a \times 10^{b}$. The value of the inverse of the fine-structure constant used in the calculations is $\alpha^{-1}=137.035 \: 999 \: 139$ (from CODATA 2014).}
\label{tab:3d_5-2_mu_1-2}
$Z$ & \multicolumn{1}{c}{$\chi_{\textrm{M}1 \to \textrm{E}2}[a_0^4]$} & $Z$ & \multicolumn{1}{c}{$\chi_{\textrm{M}1 \to \textrm{E}2}[a_0^4]$} & $Z$ & \multicolumn{1}{c}{$\chi_{\textrm{M}1 \to \textrm{E}2}[a_0^4]$} \\
\hline\\
\endfirsthead
\caption[]{(continued)}
$Z$ & \multicolumn{1}{c}{$\chi_{\textrm{M}1 \to \textrm{E}2}[a_0^4]$} & $Z$ & \multicolumn{1}{c}{$\chi_{\textrm{M}1 \to \textrm{E}2}[a_0^4]$} & $Z$ & \multicolumn{1}{c}{$\chi_{\textrm{M}1 \to \textrm{E}2}[a_0^4]$} \\
\hline\\
\endhead
1    &   $-$1.918 025 346 563($+$5)    &   47   &   $-$1.697 539 718 372($-$5)   &   93    &   $-$2.437 289 718 316($-$7)   \\

2    &   $-$2.996 726 514 005($+$3)    &   48   &   $-$1.493 007 774 094($-$5)   &   94    &   $-$2.275 290 465 957($-$7)   \\

3    &   $-$2.630 595 806 337($+$2)    &   49   &   $-$1.316 482 075 014($-$5)   &   95    &   $-$2.125 358 239 743($-$7)   \\

4    &   $-$4.681 209 650 674($+$1)    &   50   &   $-$1.163 677 863 631($-$5)   &   96    &   $-$1.986 490 735 703($-$7)   \\

5    &   $-$1.226 919 914 208($+$1)    &   51   &   $-$1.031 032 690 132($-$5)   &   97    &   $-$1.857 777 646 955($-$7)   \\

6    &   $-$4.107 984 033 203($+$0)    &   52   &   $-$9.155 737 601 342($-$6)   &   98    &   $-$1.738 391 417 610($-$7)   \\

7    &   $-$1.628 657 682 754($+$0)    &   53   &   $-$8.148 115 765 022($-$6)   &   99    &   $-$1.627 579 005 859($-$7)   \\

8    &   $-$7.307 043 908 474($-$1)   &   54   &   $-$7.266 542 932 227($-$6)   &   100   &   $-$1.524 654 537 481($-$7)   \\
\hline
\end{longtable}

\newpage

\footnotesize
\begin{longtable}{@{\extracolsep\fill}rlrlrl@{}}
\caption{The magnetic-dipole-to-electric-quadrupole cross-susceptibilities $\chi_{\textrm{M}1 \to \textrm{E}2}$ for Dirac one-electron atoms in the excited state 3d$_{5/2}$ ($|\mu|=3/2$), computed from the analytical formula in Eq.\ (\ref{eq:2}). Notation $a(b)$ denotes $a \times 10^{b}$. The value of the inverse of the fine-structure constant used in the calculations is $\alpha^{-1}=137.035 \: 999 \: 139$ (from CODATA 2014).}
\label{tab:3d_5-2_mu_3-2}
$Z$ & \multicolumn{1}{c}{$\chi_{\textrm{M}1 \to \textrm{E}2}[a_0^4]$} & $Z$ & \multicolumn{1}{c}{$\chi_{\textrm{M}1 \to \textrm{E}2}[a_0^4]$} & $Z$ & \multicolumn{1}{c}{$\chi_{\textrm{M}1 \to \textrm{E}2}[a_0^4]$} \\
\hline\\
\endfirsthead
\caption[]{(continued)}
$Z$ & \multicolumn{1}{c}{$\chi_{\textrm{M}1 \to \textrm{E}2}[a_0^4]$} & $Z$ & \multicolumn{1}{c}{$\chi_{\textrm{M}1 \to \textrm{E}2}[a_0^4]$} & $Z$ & \multicolumn{1}{c}{$\chi_{\textrm{M}1 \to \textrm{E}2}[a_0^4]$} \\
\hline\\
\endhead
1    &   $-$3.836 050 693 195($+$5)    &   47   &   $-$3.395 388 877 217($-$5)   &   93    &   $-$4.882 240 261 603($-$7)   \\

2    &   $-$5.993 453 029 737($+$3)    &   48   &   $-$2.986 312 095 647($-$5)   &   94    &   $-$4.558 072 720 739($-$7)   \\

3    &   $-$5.261 191 620 348($+$2)    &   49   &   $-$2.633 248 585 749($-$5)   &   95    &   $-$4.258 044 541 093($-$7)   \\

4    &   $-$9.362 419 344 514($+$1)    &   50   &   $-$2.327 628 770 650($-$5)   &   96    &   $-$3.980 150 894 665($-$7)   \\

5    &   $-$2.453 839 856 041($+$1)    &   51   &   $-$2.062 327 694 854($-$5)   &   97    &   $-$3.722 570 958 915($-$7)   \\

6    &   $-$8.215 968 258 241($+$0)    &   52   &   $-$1.831 399 719 050($-$5)   &   98    &   $-$3.483 649 423 886($-$7)   \\

7    &   $-$3.257 315 506 440($+$0)    &   53   &   $-$1.629 865 803 143($-$5)   &   99    &   $-$3.261 880 017 741($-$7)   \\

8    &   $-$1.461 408 889 588($+$0)    &   54   &   $-$1.453 542 213 486($-$5)   &   100   &   $-$3.055 890 813 177($-$7)   \\
\hline
\end{longtable}

\newpage

\footnotesize
\begin{longtable}{@{\extracolsep\fill}rlrlrl@{}}
\caption{The magnetic-dipole-to-electric-quadrupole cross-susceptibilities $\chi_{\textrm{M}1 \to \textrm{E}2}$ for Dirac one-electron atoms in the excited state 3d$_{5/2}$ ($|\mu|=5/2$), computed from the analytical formula in Eq.\ (\ref{eq:2}). Notation $a(b)$ denotes $a \times 10^{b}$. The value of the inverse of the fine-structure constant used in the calculations is $\alpha^{-1}=137.035 \: 999 \: 139$ (from CODATA 2014).}
\label{tab:3d_5-2_mu_5-2}
$Z$ & \multicolumn{1}{c}{$\chi_{\textrm{M}1 \to \textrm{E}2}[a_0^4]$} & $Z$ & \multicolumn{1}{c}{$\chi_{\textrm{M}1 \to \textrm{E}2}[a_0^4]$} & $Z$ & \multicolumn{1}{c}{$\chi_{\textrm{M}1 \to \textrm{E}2}[a_0^4]$} \\
\hline\\
\endfirsthead
\caption[]{(continued)}
$Z$ & \multicolumn{1}{c}{$\chi_{\textrm{M}1 \to \textrm{E}2}[a_0^4]$} & $Z$ & \multicolumn{1}{c}{$\chi_{\textrm{M}1 \to \textrm{E}2}[a_0^4]$} & $Z$ & \multicolumn{1}{c}{$\chi_{\textrm{M}1 \to \textrm{E}2}[a_0^4]$} \\
\hline\\
\endhead
1    &   $-$3.453 612 138 678($-$5)   &   47   &   $-$1.547 202 365 816($-$8)   &   93    &   $-$3.830 412 485 671($-$9)   \\

2    &   $-$8.633 908 689 193($-$6)   &   48   &   $-$1.482 737 293 186($-$8)   &   94    &   $-$3.745 894 412 023($-$9)   \\

3    &   $-$3.837 202 633 847($-$6)   &   49   &   $-$1.422 178 604 311($-$8)   &   95    &   $-$3.664 030 803 376($-$9)   \\

4    &   $-$2.158 355 514 195($-$6)   &   50   &   $-$1.365 216 934 405($-$8)   &   96    &   $-$3.584 711 629 553($-$9)   \\

5    &   $-$1.381 289 132 792($-$6)   &   51   &   $-$1.311 572 948 142($-$8)   &   97    &   $-$3.507 832 502 614($-$9)   \\

6    &   $-$9.591 789 993 579($-$7)   &   52   &   $-$1.260 993 908 488($-$8)   &   98    &   $-$3.433 294 333 195($-$9)   \\

7    &   $-$7.046 599 018 639($-$7)   &   53   &   $-$1.213 250 694 362($-$8)   &   99    &   $-$3.361 003 011 012($-$9)   \\

8    &   $-$5.394 672 180 440($-$7)   &   54   &   $-$1.168 135 201 254($-$8)   &   100   &   $-$3.290 869 107 617($-$9)   \\
\hline
\end{longtable}

\newpage

\footnotesize
\begin{longtable}{@{\extracolsep\fill}rlrlrl@{}}
\caption{The magnetic-dipole-to-electric-quadrupole cross-susceptibilities $\chi_{\textrm{M}1 \to \textrm{E}2}$ for Dirac one-electron atoms in the excited state 4s$_{1/2}$, computed from the analytical formula in Eq.\ (\ref{eq:2}). Notation $a(b)$ denotes $a \times 10^{b}$. The value of the inverse of the fine-structure constant used in the calculations is $\alpha^{-1}=137.035 \: 999 \: 139$ (from CODATA 2014).}
\label{tab:4s_1-2}
$Z$ & \multicolumn{1}{c}{$\chi_{\textrm{M}1 \to \textrm{E}2}[a_0^4]$} & $Z$ & \multicolumn{1}{c}{$\chi_{\textrm{M}1 \to \textrm{E}2}[a_0^4]$} & $Z$ & \multicolumn{1}{c}{$\chi_{\textrm{M}1 \to \textrm{E}2}[a_0^4]$} \\
\hline\\
\endfirsthead
\caption[]{(continued)}
$Z$ & \multicolumn{1}{c}{$\chi_{\textrm{M}1 \to \textrm{E}2}[a_0^4]$} & $Z$ & \multicolumn{1}{c}{$\chi_{\textrm{M}1 \to \textrm{E}2}[a_0^4]$} & $Z$ & \multicolumn{1}{c}{$\chi_{\textrm{M}1 \to \textrm{E}2}[a_0^4]$} \\
\hline\\
\endhead
1    &   $-$9.340 413 868 677($-$1)   &   47   &   $-$1.814 310 799 427($-$7)   &   93    &   $-$9.152 881 689 078($-$9)                              \\

2    &   $-$5.837 388 433 303($-$2)   &   48   &   $-$1.663 397 574 553($-$7)   &   94    &   $-$8.677 793 901 650($-$9)                              \\

3    &   $-$1.152 942 441 840($-$2)   &   49   &   $-$1.527 558 468 705($-$7)   &   95    &   $-$8.228 234 522 447($-$9)                              \\

4    &   $-$3.647 441 442 372($-$3)   &   50   &   $-$1.405 030 188 502($-$7)   &   96    &   $-$7.802 596 706 570($-$9)                              \\

5    &   $-$1.493 707 127 641($-$3)   &   51   &   $-$1.294 285 002 586($-$7)   &   97    &   $-$7.399 388 579 645($-$9)                              \\

6    &   $-$7.201 767 751 084($-$4)   &   52   &   $-$1.193 995 486 981($-$7)   &   98    &   $-$7.017 223 891 718($-$9)                              \\

7    &   $-$3.886 261 171 609($-$4)   &   53   &   $-$1.103 005 093 669($-$7)   &   99    &   $-$6.654 813 506 641($-$9)                              \\

8    &   $-$2.277 327 662 425($-$4)   &   54   &   $-$1.020 303 493 346($-$7)   &   100   &   $-$6.310 957 643 631($-$9)                              \\
\hline
\end{longtable}

\newpage

\footnotesize
\begin{longtable}{@{\extracolsep\fill}rlrlrl@{}}
\caption{The magnetic-dipole-to-electric-quadrupole cross-susceptibilities $\chi_{\textrm{M}1 \to \textrm{E}2}$ for Dirac one-electron atoms in the excited state 4p$_{1/2}$, computed from the analytical formula in Eq.\ (\ref{eq:2}). Notation $a(b)$ denotes $a \times 10^{b}$. The value of the inverse of the fine-structure constant used in the calculations is $\alpha^{-1}=137.035 \: 999 \: 139$ (from CODATA 2014).}
\label{tab:4p_1-2}
$Z$ & \multicolumn{1}{c}{$\chi_{\textrm{M}1 \to \textrm{E}2}[a_0^4]$} & $Z$ & \multicolumn{1}{c}{$\chi_{\textrm{M}1 \to \textrm{E}2}[a_0^4]$} & $Z$ & \multicolumn{1}{c}{$\chi_{\textrm{M}1 \to \textrm{E}2}[a_0^4]$} \\
\hline\\
\endfirsthead
\caption[]{(continued)}
$Z$ & \multicolumn{1}{c}{$\chi_{\textrm{M}1 \to \textrm{E}2}[a_0^4]$} & $Z$ & \multicolumn{1}{c}{$\chi_{\textrm{M}1 \to \textrm{E}2}[a_0^4]$} & $Z$ & \multicolumn{1}{c}{$\chi_{\textrm{M}1 \to \textrm{E}2}[a_0^4]$} \\
\hline\\
\endhead
1    &   2.806 355 191 331($+$6)    &   47   &   2.311 107 751 685($-$4)   &   93    &   2.414 619 775 930($-$6)                              \\

2    &   4.384 264 022 130($+$4)    &   48   &   2.025 720 017 144($-$4)   &   94    &   2.225 292 306 957($-$6)                              \\

3    &   3.848 036 486 544($+$3)    &   49   &   1.779 940 448 738($-$4)   &   95    &   2.051 165 477 595($-$6)                              \\

4    &   6.846 250 109 001($+$2)    &   50   &   1.567 658 972 023($-$4)   &   96    &   1.890 922 729 351($-$6)                              \\

5    &   1.793 884 981 568($+$2)    &   51   &   1.383 801 172 094($-$4)   &   97    &   1.743 371 454 114($-$6)                              \\

6    &   6.004 334 446 795($+$1)    &   52   &   1.224 137 366 163($-$4)   &   98    &   1.607 430 317 687($-$6)                              \\

7    &   2.379 565 476 732($+$1)    &   53   &   1.085 129 626 854($-$4)   &   99    &   1.482 117 983 335($-$6)                              \\

8    &   1.067 124 693 358($+$1)    &   54   &   9.638 086 826 693($-$5)   &   100   &   1.366 543 069 059($-$6)                              \\
\hline
\end{longtable}

\newpage

\footnotesize
\begin{longtable}{@{\extracolsep\fill}rlrlrl@{}}
\caption{The magnetic-dipole-to-electric-quadrupole cross-susceptibilities $\chi_{\textrm{M}1 \to \textrm{E}2}$ for Dirac one-electron atoms in the excited state 4p$_{3/2}$ ($|\mu|=1/2$), computed from the analytical formula in Eq.\ (\ref{eq:2}). Notation $a(b)$ denotes $a \times 10^{b}$. The value of the inverse of the fine-structure constant used in the calculations is $\alpha^{-1}=137.035 \: 999 \: 139$ (from CODATA 2014).}
\label{tab:4p_3-2_mu_1-2}
$Z$ & \multicolumn{1}{c}{$\chi_{\textrm{M}1 \to \textrm{E}2}[a_0^4]$} & $Z$ & \multicolumn{1}{c}{$\chi_{\textrm{M}1 \to \textrm{E}2}[a_0^4]$} & $Z$ & \multicolumn{1}{c}{$\chi_{\textrm{M}1 \to \textrm{E}2}[a_0^4]$} \\
\hline\\
\endfirsthead
\caption[]{(continued)}
$Z$ & \multicolumn{1}{c}{$\chi_{\textrm{M}1 \to \textrm{E}2}[a_0^4]$} & $Z$ & \multicolumn{1}{c}{$\chi_{\textrm{M}1 \to \textrm{E}2}[a_0^4]$} & $Z$ & \multicolumn{1}{c}{$\chi_{\textrm{M}1 \to \textrm{E}2}[a_0^4]$} \\
\hline\\
\endhead
1    &   $-$2.806 355 431 460($+$6)    &   47   &   $-$2.310 927 638 896($-$4)   &   93    &   $-$2.360 267 696 924($-$6)                            \\

2    &   $-$4.384 265 521 853($+$4)    &   48   &   $-$2.025 509 893 310($-$4)   &   94    &   $-$2.170 537 132 195($-$6)                            \\

3    &   $-$3.848 039 445 368($+$3)    &   49   &   $-$1.779 703 564 320($-$4)   &   95    &   $-$1.995 994 447 059($-$6)                            \\

4    &   $-$6.846 259 454 862($+$2)    &   50   &   $-$1.567 398 170 510($-$4)   &   96    &   $-$1.835 321 998 550($-$6)                            \\

5    &   $-$1.793 888 801 094($+$2)    &   51   &   $-$1.383 518 945 966($-$4)   &   97    &   $-$1.687 326 047 921($-$6)                            \\

6    &   $-$6.004 352 815 591($+$1)    &   52   &   $-$1.223 835 904 321($-$4)   &   98    &   $-$1.550 924 075 222($-$6)                            \\

7    &   $-$2.379 575 358 717($+$1)    &   53   &   $-$1.084 810 854 922($-$4)   &   99    &   $-$1.425 133 493 149($-$6)                            \\

8    &   $-$1.067 130 463 252($+$1)    &   54   &   $-$9.634 742 973 598($-$5)   &   100   &   $-$1.309 061 593 680($-$6)                            \\
\hline
\end{longtable}

\newpage

\footnotesize
\begin{longtable}{@{\extracolsep\fill}rlrlrl@{}}
\caption{The magnetic-dipole-to-electric-quadrupole cross-susceptibilities $\chi_{\textrm{M}1 \to \textrm{E}2}$ for Dirac one-electron atoms in the excited state 4p$_{3/2}$ ($|\mu|=3/2$), computed from the analytical formula in Eq.\ (\ref{eq:2}). Notation $a(b)$ denotes $a \times 10^{b}$. The value of the inverse of the fine-structure constant used in the calculations is $\alpha^{-1}=137.035 \: 999 \: 139$ (from CODATA 2014).}
\label{tab:4p_3-2_mu_3-2}
$Z$ & \multicolumn{1}{c}{$\chi_{\textrm{M}1 \to \textrm{E}2}[a_0^4]$} & $Z$ & \multicolumn{1}{c}{$\chi_{\textrm{M}1 \to \textrm{E}2}[a_0^4]$} & $Z$ & \multicolumn{1}{c}{$\chi_{\textrm{M}1 \to \textrm{E}2}[a_0^4]$} \\
\hline\\
\endfirsthead
\caption[]{(continued)}
$Z$ & \multicolumn{1}{c}{$\chi_{\textrm{M}1 \to \textrm{E}2}[a_0^4]$} & $Z$ & \multicolumn{1}{c}{$\chi_{\textrm{M}1 \to \textrm{E}2}[a_0^4]$} & $Z$ & \multicolumn{1}{c}{$\chi_{\textrm{M}1 \to \textrm{E}2}[a_0^4]$} \\
\hline\\
\endhead
1    &   $-$4.804 454 284 160($-$1)   &   47   &   $-$1.298 358 204 540($-$7)   &   93    &   $-$1.383 893 694 220($-$8)   \\

2    &   $-$3.004 117 624 806($-$2)   &   48   &   $-$1.205 572 561 503($-$7)   &   94    &   $-$1.339 602 078 671($-$8)   \\

3    &   $-$5.938 449 877 921($-$3)   &   49   &   $-$1.121 458 766 669($-$7)   &   95    &   $-$1.297 250 493 065($-$8)   \\

4    &   $-$1.880 907 188 189($-$3)   &   50   &   $-$1.045 037 170 578($-$7)   &   96    &   $-$1.256 730 554 247($-$8)   \\

5    &   $-$7.714 434 432 473($-$4)   &   51   &   $-$9.754 563 379 409($-$8)   &   97    &   $-$1.217 941 131 606($-$8)   \\

6    &   $-$3.726 343 272 167($-$4)   &   52   &   $-$9.119 741 091 019($-$8)   &   98    &   $-$1.180 787 786 724($-$8)   \\

7    &   $-$2.015 235 073 905($-$4)   &   53   &   $-$8.539 417 619 376($-$8)   &   99    &   $-$1.145 182 261 832($-$8)   \\

8    &   $-$1.183 895 471 245($-$4)   &   54   &   $-$8.007 907 191 842($-$8)   &   100   &   $-$1.111 042 012 364($-$8)   \\
\hline
\end{longtable}

\newpage

\footnotesize
\begin{longtable}{@{\extracolsep\fill}rlrlrl@{}}
\caption{The magnetic-dipole-to-electric-quadrupole cross-susceptibilities $\chi_{\textrm{M}1 \to \textrm{E}2}$ for Dirac one-electron atoms in the excited state 4d$_{3/2}$ ($|\mu|=1/2$), computed from the analytical formula in Eq.\ (\ref{eq:2}). Notation $a(b)$ denotes $a \times 10^{b}$. The value of the inverse of the fine-structure constant used in the calculations is $\alpha^{-1}=137.035 \: 999 \: 139$ (from CODATA 2014).}
\label{tab:4d_3-2_mu_1-2}
$Z$ & \multicolumn{1}{c}{$\chi_{\textrm{M}1 \to \textrm{E}2}[a_0^4]$} & $Z$ & \multicolumn{1}{c}{$\chi_{\textrm{M}1 \to \textrm{E}2}[a_0^4]$} & $Z$ & \multicolumn{1}{c}{$\chi_{\textrm{M}1 \to \textrm{E}2}[a_0^4]$} \\
\hline\\
\endfirsthead
\caption[]{(continued)}
$Z$ & \multicolumn{1}{c}{$\chi_{\textrm{M}1 \to \textrm{E}2}[a_0^4]$} & $Z$ & \multicolumn{1}{c}{$\chi_{\textrm{M}1 \to \textrm{E}2}[a_0^4]$} & $Z$ & \multicolumn{1}{c}{$\chi_{\textrm{M}1 \to \textrm{E}2}[a_0^4]$} \\
\hline\\
\endhead
1    &   1.818 576 084 845($+$6)    &   47   &   1.617 627 975 841($-$4)   &   93    &   2.365 583 375 893($-$6)                          \\

2    &   2.841 365 097 716($+$4)    &   48   &   1.423 050 617 006($-$4)   &   94    &   2.209 735 338 054($-$6)                          \\

3    &   2.494 242 747 175($+$3)    &   49   &   1.255 091 545 860($-$4)   &   95    &   2.065 443 593 491($-$6)                          \\

4    &   4.438 632 791 254($+$2)    &   50   &   1.109 680 711 239($-$4)   &   96    &   1.931 750 756 814($-$6)                          \\

5    &   1.163 364 324 322($+$2)    &   51   &   9.834 339 106 137($-$5)   &   97    &   1.807 787 167 854($-$6)                          \\

6    &   3.895 278 784 816($+$1)    &   52   &   8.735 268 676 755($-$5)   &   98    &   1.692 762 084 893($-$6)                          \\

7    &   1.544 371 430 065($+$1)    &   53   &   7.775 942 650 729($-$5)   &   99    &   1.585 955 838 320($-$6)                          \\

8    &   6.929 113 746 908($+$0)    &   54   &   6.936 484 353 259($-$5)   &   100   &   1.486 712 831 756($-$6)                          \\
\hline
\end{longtable}

\newpage
\footnotesize
\begin{longtable}{@{\extracolsep\fill}rlrlrl@{}}
\caption{The magnetic-dipole-to-electric-quadrupole cross-susceptibilities $\chi_{\textrm{M}1 \to \textrm{E}2}$ for Dirac one-electron atoms in the excited state 4d$_{3/2}$ ($|\mu|=3/2$), computed from the analytical formula in Eq.\ (\ref{eq:2}). Notation $a(b)$ denotes $a \times 10^{b}$. The value of the inverse of the fine-structure constant used in the calculations is $\alpha^{-1}=137.035 \: 999 \: 139$ (from CODATA 2014).}
\label{tab:4d_3-2_mu_3-2}
$Z$ & \multicolumn{1}{c}{$\chi_{\textrm{M}1 \to \textrm{E}2}[a_0^4]$} & $Z$ & \multicolumn{1}{c}{$\chi_{\textrm{M}1 \to \textrm{E}2}[a_0^4]$} & $Z$ & \multicolumn{1}{c}{$\chi_{\textrm{M}1 \to \textrm{E}2}[a_0^4]$} \\
\hline\\
\endfirsthead
\caption[]{(continued)}
$Z$ & \multicolumn{1}{c}{$\chi_{\textrm{M}1 \to \textrm{E}2}[a_0^4]$} & $Z$ & \multicolumn{1}{c}{$\chi_{\textrm{M}1 \to \textrm{E}2}[a_0^4]$} & $Z$ & \multicolumn{1}{c}{$\chi_{\textrm{M}1 \to \textrm{E}2}[a_0^4]$} \\
\hline\\
\endhead
1    &   3.637 150 301 476($+$6)    &   47   &   3.231 038 782 032($-$4)   &   93    &   4.698 406 395 618($-$6)   \\

2    &   5.682 718 517 375($+$4)    &   48   &   2.842 210 399 285($-$4)   &   94    &   4.387 988 368 602($-$6)   \\

3    &   4.988 462 420 925($+$3)    &   49   &   2.506 586 960 361($-$4)   &   95    &   4.100 622 001 320($-$6)   \\

4    &   8.877 192 551 691($+$2)    &   50   &   2.216 032 007 043($-$4)   &   96    &   3.834 396 165 238($-$6)   \\

5    &   2.326 698 722 030($+$2)    &   51   &   1.963 780 295 854($-$4)   &   97    &   3.587 574 931 718($-$6)   \\

6    &   7.790 413 169 911($+$1)    &   52   &   1.744 186 025 036($-$4)   &   98    &   3.358 579 978 570($-$6)   \\

7    &   3.088 664 867 221($+$1)    &   53   &   1.552 520 960 030($-$4)   &   99    &   3.145 974 915 693($-$6)   \\

8    &   1.385 776 997 894($+$1)    &   54   &   1.384 811 865 593($-$4)   &   100   &   2.948 451 304 086($-$6)   \\
\hline
\end{longtable}

\newpage

\footnotesize
\begin{longtable}{@{\extracolsep\fill}rlrlrl@{}}
\caption{The magnetic-dipole-to-electric-quadrupole cross-susceptibilities $\chi_{\textrm{M}1 \to \textrm{E}2}$ for Dirac one-electron atoms in the excited state 4d$_{5/2}$ ($|\mu|=1/2$), computed from the analytical formula in Eq.\ (\ref{eq:2}). Notation $a(b)$ denotes $a \times 10^{b}$. The value of the inverse of the fine-structure constant used in the calculations is $\alpha^{-1}=137.035 \: 999 \: 139$ (from CODATA 2014).}
\label{tab:4d_5-2_mu_1-2}
$Z$ & \multicolumn{1}{c}{$\chi_{\textrm{M}1 \to \textrm{E}2}[a_0^4]$} & $Z$ & \multicolumn{1}{c}{$\chi_{\textrm{M}1 \to \textrm{E}2}[a_0^4]$} & $Z$ & \multicolumn{1}{c}{$\chi_{\textrm{M}1 \to \textrm{E}2}[a_0^4]$} \\
\hline\\
\endfirsthead
\caption[]{(continued)}
$Z$ & \multicolumn{1}{c}{$\chi_{\textrm{M}1 \to \textrm{E}2}[a_0^4]$} & $Z$ & \multicolumn{1}{c}{$\chi_{\textrm{M}1 \to \textrm{E}2}[a_0^4]$} & $Z$ & \multicolumn{1}{c}{$\chi_{\textrm{M}1 \to \textrm{E}2}[a_0^4]$} \\
\hline\\
\endhead
1    &   $-$1.818 575 150 739($+$6)    &   47   &   $-$1.615 512 315 878($-$4)   &   93    &   $-$2.348 211 593 121($-$6)   \\

2    &   $-$2.841 359 258 707($+$4)    &   48   &   $-$1.421 097 975 433($-$4)   &   94    &   $-$2.193 000 028 743($-$6)   \\

3    &   $-$2.494 231 210 549($+$3)    &   49   &   $-$1.253 286 115 711($-$4)   &   95    &   $-$2.049 314 343 957($-$6)   \\

4    &   $-$4.438 596 276 325($+$2)    &   50   &   $-$1.108 008 506 900($-$4)   &   96    &   $-$1.916 198 971 841($-$6)   \\

5    &   $-$1.163 349 361 318($+$2)    &   51   &   $-$9.818 825 266 258($-$5)   &   97    &   $-$1.792 785 945 549($-$6)   \\

6    &   $-$3.895 206 587 029($+$1)    &   52   &   $-$8.720 852 734 365($-$5)   &   98    &   $-$1.678 286 101 574($-$6)   \\

7    &   $-$1.544 332 435 106($+$1)    &   53   &   $-$7.762 526 295 296($-$5)   &   99    &   $-$1.571 981 242 565($-$6)   \\

8    &   $-$6.928 885 000 672($+$0)    &   54   &   $-$6.923 979 768 119($-$5)   &   100   &   $-$1.473 217 146 820($-$6)   \\
\hline
\end{longtable}

\newpage
\footnotesize
\begin{longtable}{@{\extracolsep\fill}rlrlrl@{}}
\caption{The magnetic-dipole-to-electric-quadrupole cross-susceptibilities $\chi_{\textrm{M}1 \to \textrm{E}2}$ for Dirac one-electron atoms in the excited state 4d$_{5/2}$ ($|\mu|=3/2$), computed from the analytical formula in Eq.\ (\ref{eq:2}). Notation $a(b)$ denotes $a \times 10^{b}$. The value of the inverse of the fine-structure constant used in the calculations is $\alpha^{-1}=137.035 \: 999 \: 139$ (from CODATA 2014).}
\label{tab:4d_5-2_mu_3-2}
$Z$ & \multicolumn{1}{c}{$\chi_{\textrm{M}1 \to \textrm{E}2}[a_0^4]$} & $Z$ & \multicolumn{1}{c}{$\chi_{\textrm{M}1 \to \textrm{E}2}[a_0^4]$} & $Z$ & \multicolumn{1}{c}{$\chi_{\textrm{M}1 \to \textrm{E}2}[a_0^4]$} \\
\hline\\
\endfirsthead
\caption[]{(continued)}
$Z$ & \multicolumn{1}{c}{$\chi_{\textrm{M}1 \to \textrm{E}2}[a_0^4]$} & $Z$ & \multicolumn{1}{c}{$\chi_{\textrm{M}1 \to \textrm{E}2}[a_0^4]$} & $Z$ & \multicolumn{1}{c}{$\chi_{\textrm{M}1 \to \textrm{E}2}[a_0^4]$} \\
\hline\\
\endhead
1    &   $-$3.637 150 301 506($+$6)    &   47   &   $-$3.231 151 570 205($-$4)   &   93    &   $-$4.699 554 232 738($-$6)   \\

2    &   $-$5.682 718 518 123($+$4)    &   48   &   $-$2.842 317 593 693($-$4)   &   94    &   $-$4.389 061 691 629($-$6)   \\

3    &   $-$4.988 462 424 249($+$3)    &   49   &   $-$2.506 688 899 532($-$4)   &   95    &   $-$4.101 623 090 505($-$6)   \\

4    &   $-$8.877 192 570 381($+$2)    &   50   &   $-$2.216 129 002 685($-$4)   &   96    &   $-$3.835 327 205 138($-$6)   \\

5    &   $-$2.326 698 733 983($+$2)    &   51   &   $-$1.963 872 635 458($-$4)   &   97    &   $-$3.588 438 016 083($-$6)   \\

6    &   $-$7.790 413 252 852($+$1)    &   52   &   $-$1.744 273 974 183($-$4)   &   98    &   $-$3.359 377 114 924($-$6)   \\

7    &   $-$3.088 664 928 097($+$1)    &   53   &   $-$1.552 604 764 433($-$4)   &   99    &   $-$3.146 708 029 650($-$6)   \\

8    &   $-$1.385 777 044 449($+$1)    &   54   &   $-$1.384 891 752 927($-$4)   &   100   &   $-$2.949 122 243 394($-$6)   \\
\hline
\end{longtable}%

\end{document}